\documentclass[12pt]{article}

\usepackage[dvips]{graphicx}

\newcommand{\lsim}{\raisebox{-4pt}{$\,\stackrel{\textstyle
                                                         <}{\sim}\,$}}

\newcommand{\nn}{\nonumber}
\newcommand{\be}{\begin{equation}}
\newcommand{\ee}{\end{equation}}
\newcommand{\ba}{\begin{eqnarray}}
\newcommand{\ea}{\end{eqnarray}}
\newcommand{\req}[1]{(\ref{#1})}
\def\={\,=\,}
\newcommand{\ci}[1]{\cite{#1}}

\def\mev{~{\rm MeV}}
\def\gev{~{\rm GeV}}

\def\als{\alpha_{\rm s}}
\def\eps{\epsilon}

\def\muF{\mu_F}
\def\muR{\mu_R}
\def\muO{\mu_0}

\newcommand{\tw}{\textwidth}
\def\vk{{\bf k}_{\perp}}

\def\vb0{{\bf b}_0}

\newcommand{\da}{{distribution amplitude}}

\newcommand{\ov}[1]{\overline#1}

\def\={\,=\,}
\def\etap{\eta^\prime}
\def\qbq{q\bar{q}}
\def\cbc{c\bar{c}}
\def\ubu{u\bar{u}}
\def\dbd{d\bar{d}}
\def\sbs{s\bar{s}}

\usepackage{amssymb}
\newcounter{comment}

{\refstepcounter{comment}%
\begin{quote}
\ttfamily\small$\blacksquare$ \textbf{\underline{Comment} $\sharp$\thecomment:}}%
{\end{quote}}

{
\begin{quote}
\ttfamily\small$\blacktriangleright$ \textbf{\underline{Reply} $\sharp$\thecomment:}}%
{\end{quote}}

\begin{document} 
\thispagestyle{empty}
\begin{flushright}
WU B 12-01 \\
\end{flushright}

\begin{center}
{\Large\bf 
The $\eta$ ($\etap$) gamma transition form factor and
the gluon-gluon distribution amplitude
} \\
\vskip 10mm

P.\ Kroll \footnote{Email:  kroll@physik.uni-wuppertal.de}
\\[1em]
{\small {\it Fachbereich Physik, Universit\"at Wuppertal, D-42097 Wuppertal,
Germany}}\\
and\\
{\small {\it Institut f\"ur Theoretische Physik, Universit\"at
    Regensburg, \\D-93040 Regensburg, Germany}}\\
\vskip 5mm

K.\ Passek-Kumeri\v{c}ki 
\footnote{Email: passek@irb.hr}\\[1em]
{\small \it Theoretical Physics Division, Rudjer Bo\v{s}kovi\'{c} Institute, 
Zagreb, Croatia}\\

\end{center}
\vskip 5mm 
\begin{abstract}
The $\eta\gamma$ and $\etap\gamma$ transition form factors are analyzed to
leading-twist accuracy and next-to-leading order (NLO) of perturbative QCD.
Using an $\eta-\etap$ mixing scheme and all currently available experimental
data the lowest Gegenbauer coefficients of the \da s for the valence octet 
and singlet $\qbq$ and the gluon-gluon Fock components are extracted. 
Predictions for the $g^*g^*\etap$ vertex function are presented.
We also comment on the new BELLE results for the $\pi \gamma$ transition 
form factor.
\end{abstract}


\section{Introduction}

The recent measurements of the photon to pseudoscalar meson transition form
factors at large photon virtualities, $Q^2$, by the BaBar collaboration
\ci{babar09,babar} caused much excitement and renewed the interest in the
theoretical description of these observables. Most surprising is the seemingly
sharp rise of the $\pi\gamma$ form factor with the photon virtuality, which is
difficult to accommodate in fixed-order perturbative QCD. Power corrections to 
the usual leading-twist (collinear) approach \ci{BL79} seem to be required. 
However, there is also a measurement of this form factor by the BELLE
collaboration \ci{BELLE12} which shed doubts on the BaBar data. At large $Q^2$
the results of the two measurements differ, the BELLE results are close to the
theoretical expectations from the leading-twist approach. Also the 
$\eta\gamma$ and $\eta^\prime\gamma$ form factor data behave as expected 
according to theoretical analysis \ci{FK98,gluon,agaev03} of the CLEO 
\ci{cleo} and L3 data \ci{l3}, measured at lower $Q^2$ than the BaBar data. 
Strong power corrections are not demanded in these analysis. Indeed a 
next-to-leading order (NLO) leading-twist analysis \ci{gluon} is in 
reasonable agreement with the CLEO and L3 data. It is therefore 
tempting to reanalyze the $\eta\gamma$ and $\eta^\prime\gamma$ form factors to 
this accuracy along the lines presented in \ci{gluon}, taking into account the
new BaBar data. It should be mentioned that the combined CLEO, L3 and BaBar data
on these two form factors have already been analyzed in other approaches. 
Thus, for instance, in \ci{piga11} $\vk$ factorization is exploited,
and in \cite{Huang:2006wt,Wu:2011gf} in addition the non-valence 
quark contributions have been analyzed.
In \ci{Brodsky:2011xx} light-front holographic QCD was used,
in \ci{Dorokhov:2011zf} the 
low $Q^2$ data have been studied within the non-local chiral quark model and in 
\ci{Klopot:2011qq} the dispersive representation of the axial anomaly is used 
to derive an expression for the form factors that holds at all $Q^2$. 
In \ci{stech} the anomaly sum rule has 
been used for the analysis of
the transition form factors. 
A combined analysis of the low and high $Q^2$ data 
has also been performed in \ci{scopetta:11}.  

In this work we analyze the $\eta \gamma$ and $\eta'\gamma$ 
transition form factor data within the rigorously
proven collinear factorization approach. 
We restrict ourselves to the region of fairly large $Q^2$, 
and assume that in this region of $Q^2$ 
higher-twist and other power corrections are negligible. 
Thus, we have to deal with only a small number of free parameters
 
\begin{figure}
\begin{center}
\includegraphics[width=0.30\tw]{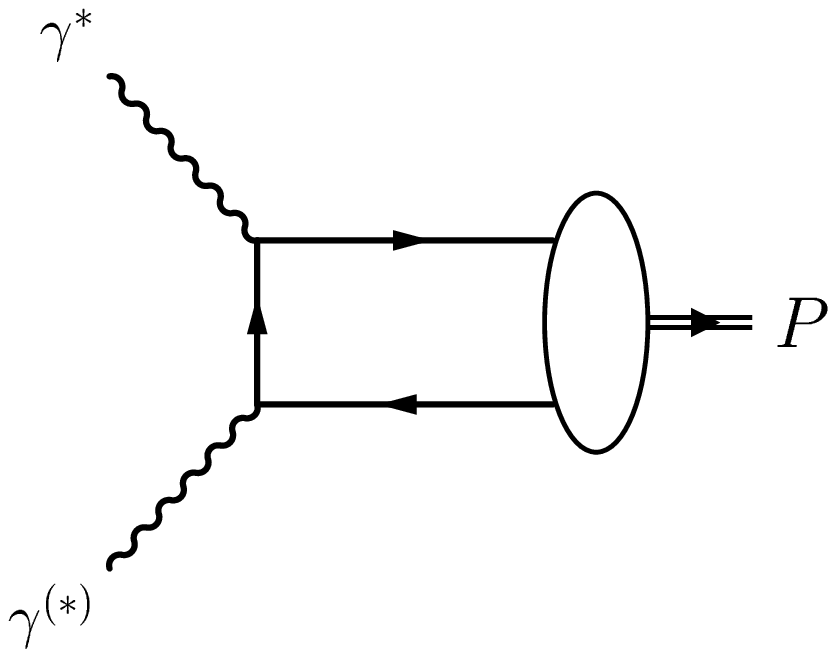}
\hspace*{0.04\tw}
\includegraphics[width=0.30\tw]{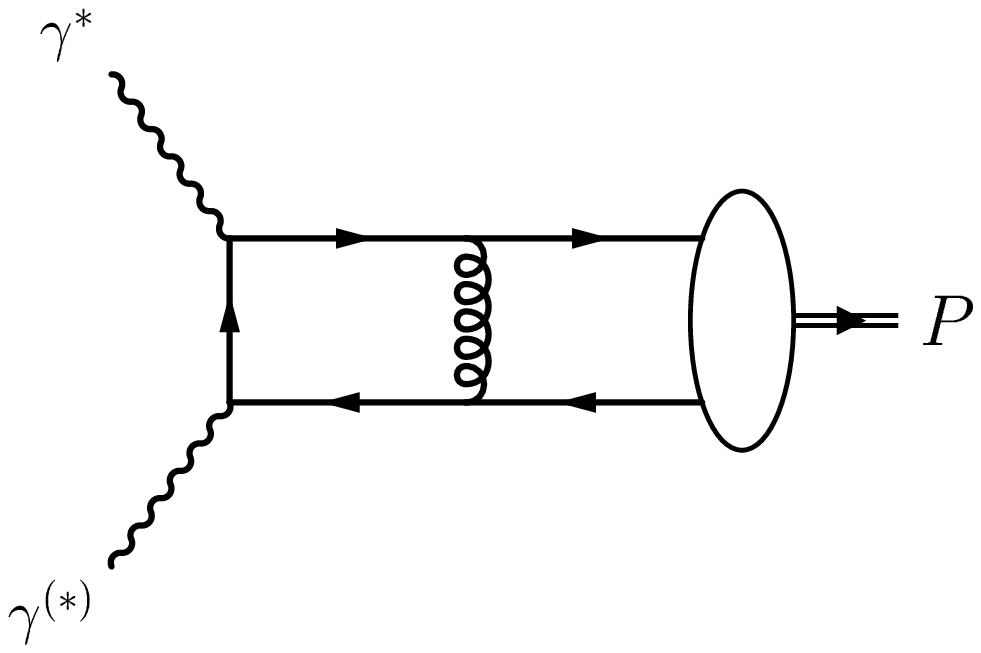}
\hspace*{-0.03\tw}
\includegraphics[width=0.30\tw]{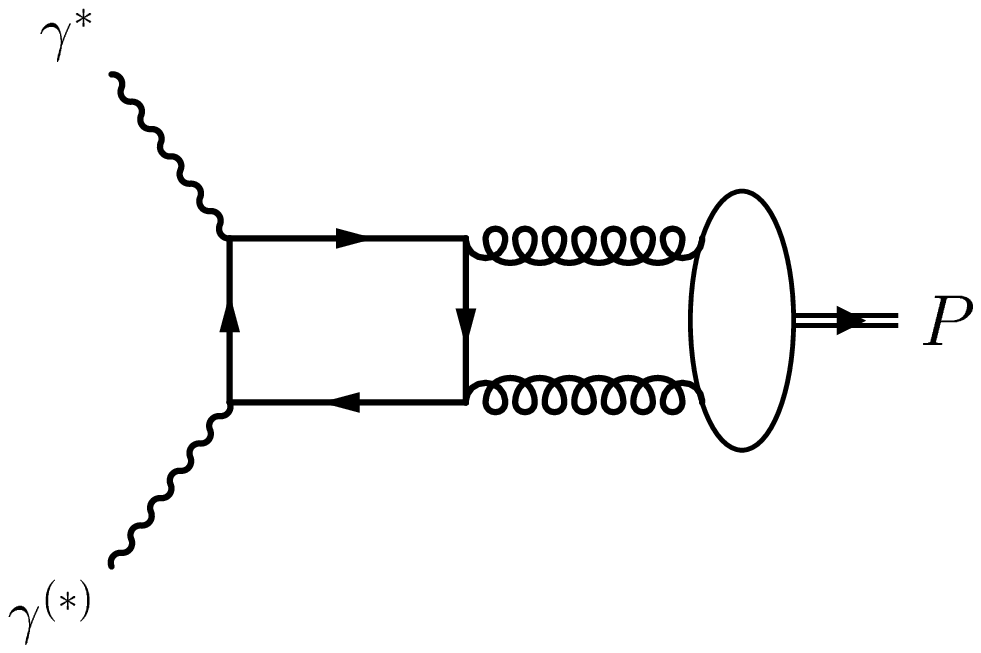} 
\end{center}
\caption{Sample Feynman graphs contribution to the transition form factors to
  NLO.}
\label{fig:feynman}
\end{figure}

The $\eta$ and $\eta^\prime$ mesons possess $SU(3)_F$ singlet and octet
quark-antiquark Fock components and, additionally, two-gluon ones leaving
aside higher Fock states. Leading-twist distribution amplitudes, $\phi$, are
associated with each of these Fock components. This feature leads, on the one
hand, to the well-known flavor mixing and, on the other hand, as a further
complication, to mixing of the $q\bar{q}$ singlet and the gluon-gluon ($gg$) 
distribution amplitudes under evolution. 
In the case of the transition form factors the two-gluon Fock components 
do not contribute to leading order (LO),
they require higher orders of perturbative QCD 
(see Fig.\ \ref{fig:feynman} for relevant Feynman graphs). 
In our previous analysis 
\ci{gluon} of the $\eta\gamma$ and $\eta^\prime\gamma$ transition form factors 
to NLO leading-twist accuracy 
the short lever arm provided by the $Q^2$
range of the CLEO and L3 data, the overall number of the data as well as the 
size of their errors, made it impossible to fix precisely even the lowest 
Gegenbauer coefficients of the distribution amplitudes. 
Here, in this work we attempt a reanalysis of the form factors, 
using in addition to the CLEO and L3 data also the recent BaBar 
data \ci{babar}. 
With our analysis we want to demonstrate that 
the $\eta$ and $\etap\gamma$ form factor data can easily be accommodated 
by a QCD calculation to NLO leading-twist accuracy. 
And, on the strength of a larger set of data covering a substantially 
larger range of $Q^2$, 
a second goal of our analysis is 
better determination of the $gg$ \da{}s 
and the discussion of accompanying theoretical uncertainties. 
This is of utmost importance since the two-gluon Fock components 
play a role in many hard 
exclusive processes involving $\eta$ and/or $\etap$ mesons which are nowadays 
accessible to experiment. Thus, for instance, the $g^*g^*\eta (\etap)$
vertex contributes to decay processes such as $\Upsilon(1S)\to\etap X$ 
(see e.g.\ \ci{ali03,ali2-03}), to the hadronic production 
$pp(\bar{p})\to \etap X$ and to meson pair production in the central 
region of proton-proton collisions \ci{khoze11}. The two-gluon Fock components
may also matter in $\chi_{cJ}$ decays in pairs of $\eta$ or $\eta^\prime$ mesons 
and they may be attributed to the enhancement of some of the $\eta^\prime$
channels in charmless $B$ decays as compared to the corresponding pion
channels (e.g.\ \ci{gronau,beneke03,williamson06}). An example is set
by the $\etap K$ channels for which such an enhancement is
experimentally observed \ci{belle06,babar2-09}. In this context the 
$B\to\eta (\etap)$ form factors are of importance which are also affected by
the $gg$ component of the $\eta$ and $\etap$ mesons (e.g. \ci{ball07}).
 
The plan of the paper is the following: 
In Sect.\ 2 we recapitulate properties of the $q\bar{q}$ and $gg$ \da s 
for the $\eta$ and $\etap$ mesons, in particular their evolution behavior. 
In Sect.\ 3 the $\eta-\etap$ mixing is briefly discussed. 
The CLEO \ci{cleo} and BaBar \ci{babar} data on the transition 
form factors are analyzed for several scenarios in Sect.\ 4 
and values for the second Gegenbauer coefficients of 
the three \da s are extracted and discussed. 
We also shortly comment on the data for the time-like transition 
form factors obtained by BaBar \ci{babar06} and 
implications of the resulting \da s for the $g^*g^*P$ vertex. 
A brief comment on the $\pi\gamma$ transition from factor is 
given in Sect.\ 5. As usual the paper ends with a summary. 

\section{Properties of the \da s}
In this work we follow the definitions and convention used in \cite{gluon}.
For the convenience of the reader we summarize here the main ingredients 
necessary for understanding and used for obtaining the results of this work.

When considering the evolution and mixing of quark and gluon states
under evolution, 
it is convenient to choose 
as valence Fock components of the pseudoscalar mesons 
the $SU(3)_F$ octet 
$|\qbq_8\rangle = |(\ubu+\dbd-2\sbs)/\sqrt{6}\rangle$
and singlet
$|\qbq_1\rangle = |(\ubu+\dbd+\sbs)/\sqrt{3}\rangle$ 
combinations of quark-antiquark states 
and the two-gluon state, $|gg\rangle$, which also possess flavor-singlet 
quantum numbers and contributes to leading-twist order. 
Higher Fock components are neglected in our analysis since 
their contributions are power
suppressed. 
It is to be stressed that the above states are partonic Fock components and
not effective meson states or glueballs which are frequently considered in the 
treatment of $\eta$ - $\etap$ mixing. 

In collinear approximation and to leading-twist accuracy a \da{}, $\phi_{Pi}$
($i=1,8,g$, $P=\eta, \etap$),
is associated with each of the Fock components we consider~\footnote{
A formal definition of the leading-twist \da s in terms of particle-vacuum
matrix elements of quark field operators or the gluon field strength tensor
can be found in \ci{gluon}.}.
These quark \da s are symmetric in $x \to (1-x)$
and normalized to unity at any factorization scale $\muF$,
while the gluon \da{} is antisymmetric 
and, consequently, there is no natural way to normalize it.
The mixing of gluon \da\ with the quark \da\ under evolution 
removes this ambiguity and, as shown in \cite{gluon}, 
the change in gluon normalization is 
reflected in the change of off-diagonal anomalous dimensions governing 
the evolution.
In \ci{gluon} an attempt was made to give 
a detailed account of different conventions
found in defining the gluon \da{} throughout the literature%

The \da s can be expanded upon the Gegenbauer polynomials
\ba
\phi_{Pi}(x,\muF)  &=& 6 x\, (1-x)\, \Big[ 1 
     + \sum_{n=2, 4, \ldots}
    {\  a_{Pn}^{\,i}(\muF)} \:  C_n^{\,3/2}(2 x -1) \Big]\,,
                                              \nn\\[0.2cm] 
\phi_{Pg}(x,\muF)  &=&  x^2 (1-x)^2 \sum_{n=2, 4, \ldots}
              { a_{Pn}^{\,g}(\muF)} \:  C_{n-1}^{\,5/2}(2 x -1) \, ,
\label{eq:solphi} 
\ea
where only the terms for even $n$ occur as a consequence of the symmetry
relations. In terms of the expansion coefficients $a_{Pn}$ 
the mixing of the quark singlet and gluon \da s are expressed by
\ba
{ a_{Pn}^{\,1}\,(\muF)} & = & a_{Pn}^{(+)}(\mu_0) 
       \left( \frac{\als(\mu_0)}{\als(\muF)} \right)^{{ \gamma^{\,(+)}_n}/{\beta_0}} 
     \hspace*{-0.02\tw}       + { \rho_n^{\,(-)}} \,  a_{Pn}^{(-)}\,(\mu_0) 
      \left( \frac{\als(\mu_0)}{\als(\muF)} \right)^{{\gamma^{\,(-)}_n}/{\beta_0}}
                   \hspace*{-0.02\tw} , \nn\\[0.2em]
{ a_{Pn}^{\,g}\,(\muF)} & = & {\rho_n^{\,(+)}} \, 
               a_{Pn}^{(+)}\,(\mu_0) 
          \left( \frac{\als(\mu_0)}{\als(\muF)}\right)^{{\gamma^{\,(+)}_n}/{\beta_0}} 
         \hspace*{-0.02\tw}    + a_{Pn}^{(-)}\,(\mu_0) 
        \left( \frac{\als(\mu_0)}{\als(\muF)}
        \right)^{{\gamma^{\,(-)}_n}/{\beta_0}}
             \hspace*{-0.02\tw}, 
\label{eq:ang}
\ea
where $\mu_0$ is the initial scale of the evolution and $\beta_0=11/3N_c-2/3n_f$.
The number of colors is denoted by $N_c$ and $n_f$ is the number of active
flavors at the characteristic scale of the process. The coefficients of the 
eigenfunctions of the
matrix evolution equation which introduces mixing between quark and gluon 
distribution amplitudes, are here denoted by  $a_{Pn}^{(\pm)}$. The powers 
$\gamma_n^{(\pm)}$ are the eigenvalues of the matrix of anomalous dimensions 
\ci{Terentev81,BaierG81}
\be
\gamma^{\,(\pm)}_n = \frac{1}{2} \left[ \gamma^{qq}_n+ \gamma^{gg}_n 
        \pm \sqrt{ \left( \gamma^{qq}_n- \gamma^{gg}_n \right)^2
        +4 \gamma^{qg}_n \gamma^{gq}_n } \right] \,,
\label{eq:gamma+-}
\ee
where the LO elements read in our convention
\ba
\gamma^{\,qq}_n &=& C_F \left[
   3 + \frac{2}{(n+1)(n+2)} - 4 \sum_{i=1}^{n+1} \frac{1}{i} \right]
                  \, , \nn \\[0.3em]
\gamma^{\,qg}_n &=& C_F \; 
              \frac{n (n+3)}{3 (n+1) (n+2)} \qquad n\ge2
                        \, , \nn \\[0.3em]
\gamma^{\,gq}_n &=& N_f  \;  
                     \frac{12}{\phantom{3} (n+1) (n+2)} \qquad n\ge2
                      \, , \nn \\[0.3em]
\gamma^{\,gg}_n &=& \beta_0 + N_c \left[ \frac{8}{(n+1)(n+2)} 
             - 4 \sum_{i=1}^{n+1} \frac{1}{i} \right] \qquad n\ge2
                       \, ,  
\label{eq:andim}
\ea
with $C_F=(N_c^2-1)/(2N_c)$.
Note that we introduce the 
distinction between the quantity $N_f$ ($=3$) which counts the (fixed valence)
quark content of the meson's flavor-singlet combination and the above defined 
$n_f$, the number of active flavors at some scale, which appears in the 
$\beta$ functions and, as such, is connected to the running of the coupling 
constant. The parameters $\rho_n^{(\pm)}$ in \req{eq:ang} are given by 
\be
\rho_n^{(+)}\=6\frac{\gamma_n^{gq}}{\gamma_n^{(+)}-\gamma_n^{gg}}\,, \qquad
\rho_n^{(-)}\=\frac16\frac{\gamma_n^{qg}}{\gamma_n^{(-)}-\gamma_n^{qq}}\,.
\ee
The evolution of the octet \da{} is merely governed by 
$\gamma^{qq}_n$ and takes the simple form
\ba
{ a_{Pn}^{\,8}\,(\muF)} & = & a_{Pn}^{(8)}(\mu_0) 
       \left( \frac{\als(\mu_0)}{\als(\muF)} \right)^{{ \gamma_n^{qq}}/{\beta_0}} 
\,.
\label{eq:a8n}
\ea
As is well-known and can be seen from
(\ref{eq:solphi} - \ref{eq:a8n}), the quark \da s evolve into the asymptotic
form
\be
\phi_{\rm AS}\=6 x (1-x)
\ee
and the gluon one to zero for $Q^2\to\infty$. 

To the order we are working, NLO evolution of the quark \da{} should in
principle be included. To this accuracy the Gegenbauer polynomials $C_n^{3/2}$ 
are no longer eigenfunctions of the evolution kernel, i.e.\ their
coefficients $a^i_{Pn}$ do not evolve independently \ci{Muller95}. The impact
of the NLO evolution on the transition form factors is expected to be small
compared with the NLO corrections to the subprocess amplitudes
\ci{KrollR96}. Therefore we refrain from considering NLO evolution here.

The full quark and gluon \da{}s defined through hadronic matrix elements
read
$f_P^i/(2\sqrt{2N_c})\phi_{Pi}$ and 
$f_P^1/(2\sqrt{2N_c}) \phi_{Pg}$, respectively. 
The decay constants, $f_P^i$, are, as usual, defined by vacuum-meson matrix
elements of flavor singlet or octet weak axial-vector currents
\be
\langle 0\mid J^i_{\mu 5}(0) \mid P(p)\rangle \= i f_P^i p_\mu\,.
\ee
The singlet decay constants, $f_P^1$, depend on the scale but the anomalous
dimension controlling it is of order $\als^2$ \ci{leutwyler00}. In fact the 
evolution of $f_{P}^{1}$ is part of the NLO evolution of the singlet \da, it 
represents the scale dependence of its first Gegenbauer coefficient. This 
is to be contrasted to the octet \da{} for which the first Gegenbauer 
coefficient, $a_0$, is 1 at all scales. Thus, in harmony with the neglect of
NLO evolution of the \da s, the scale dependence of $f_{P}^{1}$ 
is neglected too.

\section{$\eta - \eta^\prime$ mixing}
\label{sec:mixing}

This section is devoted to the phenomenological aspects of 
$\eta - \eta^\prime$ mixing. 
The mixing scheme we are employing in this work is explained, 
numerical values listed, and 
results from the literature discussed.

As in \ci{gluon},
in order to reduce the number of independent \da s we follow \ci{FKS1} 
and assume meson independence of the \da s, i.e.\
\be
\phi_{Pi} \= \phi_i\,, \qquad \phi_{Pg} \= \phi_g
\ee
Hence, the mixing behavior of the valence Fock components 
and the particle dependence solely resides in the decay constants. 
Since in hard processes only small spatial quark-antiquark separations are 
probed, this assumption is sufficiently plausible - the decay constants 
play the role of wave functions at the origin of configuration space.
We work in octet-singlet basis 
and for the decay constants we use the general parameterization
\ci{leutwyler00,FKS1}
\ba
f_{\eta\phantom{'}}^8 &=& f_8 \cos{\theta_8}\,, 
        \qquad f_{\eta\phantom{'}}^1 = -f_1 \sin{\theta_1}\,, \nn\\
f_{\eta'}^8 &=& f_8 \sin{\theta_8}\,, 
              \qquad f_{\eta'}^1 =\phantom{-}f_1 \cos{\theta_1}\,.
\label{mix81}
\ea

In \ci{FKS1,FKS2} it has been observed that $\eta - \etap$ mixing is
particularly simple in the quark-flavor basis 
with valence Fock components
$|q\bar q\,\rangle=(u\bar{u}+d\bar{d})/\sqrt{2}$
and $|s\bar s\,\rangle$.
In this basis which is supposed to separate 
strange and non-strange contributions, the mixing behavior of the decay
constants is controlled by the angles $\varphi_q$ and $\varphi_s$, defined
analogously to \req{mix81}. It turned out from phenomenology that these angles
practically fall together 
 $\varphi_q=\varphi_s=\varphi$, i.e.
that we can write
\ba
f_{\eta\phantom{'}}^q &=& f_q \cos{\varphi}\,, \qquad
            f_{\eta\phantom{'}}^s = -f_s \sin{\varphi}\,, \nn\\
f_{\eta'}^q &=& f_q \sin{\varphi}\,, \qquad
                f_{\eta'}^s =\phantom{-}f_s \cos{\varphi}\,.
\label{mixqs}
\ea
This observation is supported by a QCD sum rule study \ci{penn}. 
A recent lattice QCD study \ci{gregory11} is also not in conflict with it. 
The occurrence 
of only one mixing angle in this basis is a consequence of the smallness of 
OZI rule violations which amount to only a few percent and can safely be 
neglected in most cases. $SU(3)_F$ symmetry, on the other hand, is broken at 
the level of $10 - 20\%$. 

But, although the, so-called, quark-flavor mixing scheme, 
which employs quark-flavor basis and one mixing angle, 
offers successful phenomenological description of $\eta - \eta'$ mixing,
the inclusion of $|gg \rangle$ states favors the octet-singlet basis.
This is due to the fact that  $|gg \rangle$ state mixes under evolution
with just $SU(3)_F$ singlet state, in contrast to more complicated
mixing with both $|q\bar q\,\rangle$ and $|s\bar s\,\rangle$ states when using
quark-flavor basis.
Thus when one considers two gluon states and evolution octet-singlet
basis is natural and we use it in this work.
Since for phenomenological insight the quark-flavor
scheme is useful and a lot of results are obtained using this 
scheme, we present the useful transformation formulas 
for the valence Fock components of the basis states
$\eta_q$ and $\eta_s$%
\ba
|\eta_q\rangle = \frac{f_q}{2\sqrt{2N_c}}\, 
               \left[\phi_q(x,\muF)\, |q\bar q\,\rangle
               +      \phi_{\rm opp}(x,\muF)\, |s\bar s\,\rangle
               + \sqrt{\frac23}\; \phi_g(x,\muF)\, |gg\rangle \right]
                                             \nn\\[0.2em]  
|\eta_s\rangle = \frac{f_s}{2\sqrt{2N_c}}\,
               \left[ \phi_{\rm opp}(x,\muF)\, |q\bar q\,\rangle 
                 +  \phi_s(x,\muF)\,|\,s\bar s\,\rangle 
              \, + \frac1{\sqrt{3}}\phi_g(x,\muF)\, |gg\rangle \right]
\label{qsfock} 
\ea
and
\be
\phi_q=\frac13 (\phi_8 + 2\phi_1)\,, \quad
\phi_s=\frac13 (2\phi_8 + \phi_1)\,, \quad 
\phi_{\rm opp} = \frac{\sqrt{2}}{3} (\phi_1 - \phi_8)\,.
\label{qsfock2} 
\ee
The new decay constants are related to $f_8$ and $f_1$ by
\be
f_q\=\sqrt{2f_1^2-f_8^2}\,, \qquad f_s\=\sqrt{2f_8^2-f_1^2}\,.
\ee
We see that in \req{qsfock} the $s\bar{s}$ ($q\bar{q}$) Fock component
appears in the $\eta_q$ ($\eta_s$) states,
i.e., even if we start with pure $|q\bar q\,\rangle$ and states
$|s\bar s\,\rangle$ the evolution produces the opposite states
due to different evolution of SU$_F$(3) octet and singlet states.

We now turn to numerical values of mixing parameters found in the literature.
Working in the quark-flavor basis and exploiting the divergences of the
axial-vector currents - which embody the axial-vector anomaly - the mixing
parameters in the quark-flavor mixing scheme can be expressed in terms of the 
masses of the physical mesons \ci{Kroll05}, e.g.\
\be
\sin{\varphi}\=
     \sqrt{\frac{(M_{\etap}^2-2M_{K^0}^2+M_{\pi^0}^2)(M_\eta^2-M_{\pi^0}^2)}
                 {2(M_{\etap}^2-M_\eta^2)(M_{K^0}^2-M_{\pi^0}^2)}}\,.
\ee
Evaluation of the mixing angle provides $\varphi=41.4^\circ$. Using for the 
decay constant $f_q$ in the quark-flavor basis the $SU(3)_F$
symmetry result $f_q=f_\pi$, one finds for the strange decay constant in
that basis
\be
f_s\=f_\pi\,\sqrt{\frac{(M_{\etap}^2-M_{\pi^0}^2)(M_{\eta}^2-M_{\pi^0}^2)}
{2(M_{\etap}^2-2M_{K^0}^2+M_{\pi^0}^2)(2M_{K^0}^2-M_{\pi^0}^2-M_{\eta}^2)}}\,.
\ee
Transforming these results to the octet-singlet basis, one obtains the results 
for the mixing parameters \req{mix81} that are quoted in Tab.\ \ref{tab:mix}. 
In \ci{FKS1,FKS2} 
the mixing parameters have been determined phenomenologically 
allowing for higher-orders flavor symmetry 
breaking effects. The results, obtained from the analysis of a number of 
processes involving $\eta$ and $\etap$ mesons, are also quoted in Tab.\ 
\ref{tab:mix}. For a discussion of uncertainties see \ci{FKS1}. There are a 
few more analysis, e.g.\ \ci{penn,escribano05,bernstein,weigel}, in which the 
decay constants \req{mix81} have been determined. In general the results are 
close to those obtained in \ci{FKS1,FKS2}, for a comparison see \ci{Kroll05}. 
The largest deviations from the mixing parameters given in \ci{FKS1,FKS2} has 
been reported in \ci{escribano05}, see Tab.\ \ref{tab:mix}. This 
phenomenological analysis
has been performed along the lines described in \ci{FKS1,FKS2}
considering however only a subset of the processes investigated therein but,
if at disposal, exploiting more recent data. In other papers only the mixing
angle $\varphi$ has been determined, 
e.g.\ \ci{Huang:2006as,DiDonato:2011kr,thomas07}. 
Within occasionally large errors the values for it ($39-42^\circ$) agree with 
the ones found in \ci{FKS1,FKS2}.  
 
\begin{table*}[t]
\renewcommand{\arraystretch}{1.4} 
\begin{center}
\begin{tabular}{| c || c ||c | c | c | c|c|}
\hline     
 & Ref.  & $f_8/f_\pi$  & $\theta_8$ & $f_1/f_\pi$ & $\theta_1$& $\varphi$ 
\\[0.2em] \hline
1  & \ci{Kroll05} & 1.19 & $-19.4^\circ$ & 1.10 & $-6.8^\circ$ & $41.4^\circ$\\[0.2em]
2  & \ci{FKS1,FKS2}    & 1.26 & $-21.2^\circ$ & 1.17 & $-9.2^\circ$& $39.3^\circ$ \\[0.2em]  
3  & \ci{escribano05} & 1.51 & $-23.8^\circ$& 1.29& $-2.4^\circ$ & $40.7^\circ$\\[0.2em]
\hline  
\end{tabular}
\end{center}
\caption{Decay constants in the singlet-octet basis and the mixing angle in
  the quark-flavor basis. The value of $\varphi$ quoted in the last line for is
  the average of $\varphi_q$ and $\varphi_s$ determined in \ci{escribano05}.}
\label{tab:mix}
\renewcommand{\arraystretch}{1.0}   
\end{table*}  

And we end the section with a comment.
Frequently mixing of the $\eta$ and $\etap$ is studied starting from three
basis states, $\eta_8, \eta_1$ ( or $\eta_q, \eta_s$) and a gluonic state
$\eta_g$. Despite this the values for the angle $\varphi$ controlling
$\eta$--$\etap$ mixing obtained in such analysis agree reasonably well with 
the above quoted ones. However, these mixing schemes rely on the existence  
of a rather light pseudoscalar glueball for which there is no evidence, see
the review \ci{review}. In any case, it would be a misinterpretation to
identify the gluon-gluon state we consider here with the $\eta_g$.
Our gluon-gluon state is a partonic Fock component of the $\eta$ and $\etap$. 

\section{Analysis of the form factor data and applications}
\label{sec:analysis}

This section is devoted to presenting 
the main numerical results of the paper, 
i.e., 
the results of the fits to experimental data,
and their discussion.
We also discuss  
other results from the literature and give examples
of application.

The $\eta\gamma$ and $\eta^\prime\gamma$ transition form factors can be
represented as a sum of flavor-octet and singlet contributions
\be
F_{P\gamma}(Q^2) = F_{P\gamma}^{\,8}(Q^2) + F_{P\gamma}^{\,1}(Q^2)\,,
\label{eq:ossum}
\ee
where the singlet one also includes the gluon part. This decomposition is 
completely general. 
In the asymptotic limit the transition
form factors become
\be
F_{P\gamma} \to \sqrt{\frac23}\frac{f_P^8+ 2\sqrt{2}f_P^1}{Q^2}
\label{eq:asymp}
\ee
independently of the choice of the factorization scale. 

In our analysis we do not take into account power corrections as for instance
may arise from higher-twist effects~\footnote{
It can be shown that there is no twist-3 correction to the transition form
factor. Possible twist-4 and twist-6 corrections have been discussed for the 
case of the $\pi\gamma$ form factor recently \ci{agaev11}.},
from meson masses or from quark transverse degrees of freedom. The latter seem
to be rather small for the form factors of interest, in fact much smaller than
for the $\pi\gamma$ transition form factor as is shown for instance in
\ci{piga11}. For further comments concerning power corrections see 
below.

Let us now turn to the extraction of the various \da s or rather their
Gegenbauer coefficients from the data on the transition form factors 
\ci{babar,cleo,l3}. 
The NLO prediction for the transition form factor
reads \cite{gluon} 
\ba
Q^2 F_{P\gamma} &=& a_{P0}^{\rm eff}(\muF) \left[1-\frac53\frac{\als(\muR)}{\pi}\right]
\nn\\
    &+& a_{P2}^{\rm eff}(\muF) \left[1+\frac53\frac{\als(\muR)}{\pi}
       \Big(\frac{59}{72}-\frac56\ln{\frac{Q^2}{\muF^2}}\Big)\right] \nn\\
    &+& a_{P4}^{\rm eff}(\muF) \left[1+\frac53\frac{\als(\muR)}{\pi}
       \Big(\frac{10487}{4500}-\frac{91}{75}\ln{\frac{Q^2}{\muF^2}}\Big)\right] \nn\\  
    &-& \frac{20}{3\sqrt{3}} \frac{\als(\muR)}{\pi}f_P^1\left[
   a_2^g(\muF)\Big(\frac{55}{1296}-\frac1{108}\ln{\frac{Q^2}{\muF^2}}\Big)\right.\nn\\
  && \left.\hspace*{0.08\tw} +\; a_4^g(\muF)\Big(\frac{581}{10125}
          -\frac7{675}\ln{\frac{Q^2}{\muF^2}}\Big)\right] + \ldots
\label{eq:FFexpansion}
\ea
where we introduced the abbreviation ($n=0,2,4,\ldots$, $a_0^i=1$)
\be
a_{Pn}^{\rm eff}(\muF)\=
         \sqrt{\frac23}\,\Big[f_P^8\, a_n^8(\muF)+2\sqrt{2} f_P^1\, a_n^1(\muF)\Big]\,.
\ee

Before we analyze the data a comment is in order: Inspection of
\req{eq:FFexpansion} reveals the familiar result that only due the
admittedly mild logarithmic $Q^2$ dependence generated by the evolution and
the NLO corrections one can in principle discriminate among the Gegenbauer
coefficients of different orders. However, even with the large range of 
$Q^2$ in which form factor data are available now, this logarithmic $Q^2$ 
dependence is in practice insufficient to allow for an extraction of more 
than one Gegenbauer coefficient for each of the \da s \ci{DKV1}. 
In view of this problem, we are forced to truncate the 
Gegenbauer series at $n=2$. To leading-twist accuracy the higher Gegenbauer 
coefficients are not suppressed as is the case for other approaches in which 
power corrections, accumulated in the soft end-point regions $x\to 0$ or 1, 
are taken into account \ci{piga11,agaev11}. Therefore the $n=2$ coefficients 
we are going to determine below, suffer from a truncation error; they are to 
be viewed as effective parameters which are contaminated by higher order 
Gegenbauer coefficients.

\begin{table*}[t]
\renewcommand{\arraystretch}{1.4} 
\begin{center}
\begin{tabular}{|c || c ||c | c | c | c|}
\hline     
Remarks & $\chi^2$  & $a_2^8$  & $a_2^1$ & $a_2^g$ 
                             &\req{eq:requirement2}\\[0.2em]\hline
default & 37.7 
& $-0.05\pm 0.02$ & $-0.12\pm 0.01$ & $19\pm 5$ & 0.03 \\[0.2em]    
just \ci{cleo} &  18.5 
 & $-0.07\pm 0.03$ & $-0.11\pm 0.03$ & $17\pm 11$ & 0.02 \\[0.2em]
just \ci{babar} & 15.1 
& $-0.05\pm 0.02$ & $-0.12\pm 0.01$ & $33\pm 9$ & 0.03 \\[0.2em] \hline
$\mu_R^2=Q^2/2$ &  36.9 
& $-0.01\pm 0.02$ & $-0.08\pm 0.01$ & $10\pm 4$ & 0.03 \\[0.2em]  
$\mu_R^2=\mu_F^2=Q^2/2$ &   37.5 
& $-0.01\pm 0.02$ & $-0.07\pm 0.01$ & $6\pm 3$ & 0.03  \\[0.2em]\hline
mixing \ci{escribano05}  &  45.0 
& $0.05\pm 0.02$ & $-0.16\pm 0.01$ & $11\pm 5$& 0.10  \\[0.2em]\hline  
$a_2^1=a_8^1$ &
  49.8 & $-0.11\pm 0.01$ &  $-0.11\pm 0.01$ & $21\pm 5$  &  0.0    \\[0.2em] 
$a_2^1=a_8^1$, mix. \ci{escribano05} & 240    
& $-0.16\pm 0.01$ & $-0.16\pm 0.01$ & $19\pm 4$  & 0.0  
\\[0.2em] \hline
\end{tabular}
\end{center}
\caption{Gegenbauer coefficients at $\muO=1\,\gev$ fitted to the data from
  \ci{babar} and \ci{cleo} for $Q^2\geq 2\,\gev^2$ (22 and 18 data points, 
  respectively). Except stated otherwise, the standard setting described 
  in the text (with  $\muF=\muR=Q$), and the mixing parameters 
  of \ci{FKS1} are used (see Tab.\ \ref{tab:mix}). Eq.\ \req{eq:requirement2}
  is probed at $\muO=1\,\gev$.}
\label{tab:fit}
\renewcommand{\arraystretch}{1.0}   
\end{table*}  

\subsection{Fits}
Except stated otherwise we employ the following specifications in our fits: 
As the minimum value of $Q^2$ used in the fits we take $2\,\gev^2$ and for the
initial scale of the evolution we choose $\muO=1\,\gev$. For $\als$ we use the 
two-loop expression with four flavors ($n_f=4$) and 
$\Lambda^{(4)}_{\overline{\rm MS}}=319\,\mev$ \ci{pdg}. For the factorization 
and renormalization scales we adopt the frequently used choice 
$\muF=\muR=Q$ which conveniently avoids the $\ln{(Q^2/\muF^2)}$ terms in 
\req{eq:FFexpansion} and, hence, a contingent resummation of these logarithms 
\ci{MelicNP01,MelicNP01a}. In Sect.\ \ref{sec:scale} we will comment on
other choices for these scales and accompanying theoretical uncertainties. 

\begin{figure}[t]
\begin{center}
\includegraphics[width=0.41\tw,bb=108 364 570 708,clip=true]{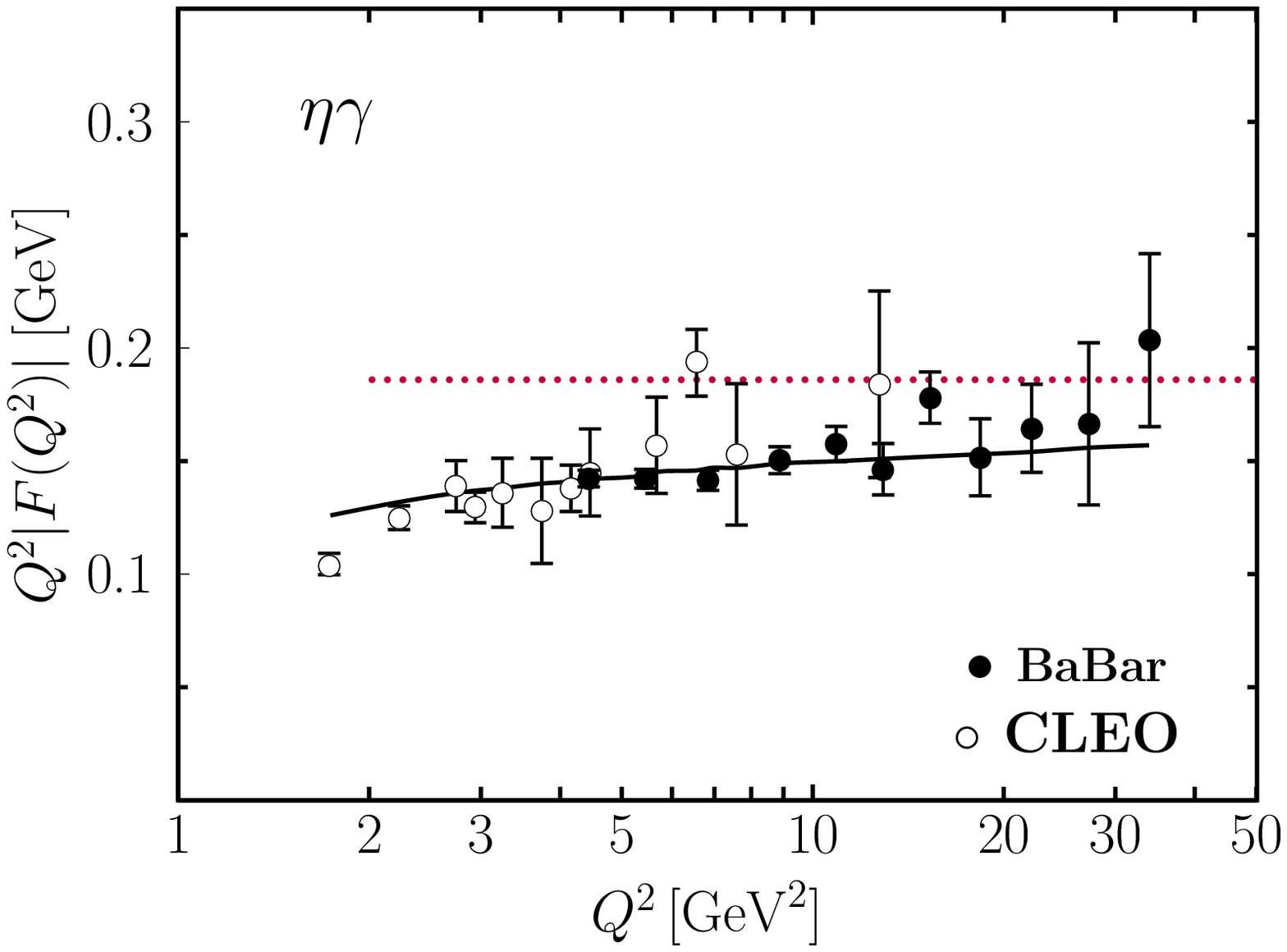}
\hspace*{0.04\tw}
\includegraphics[width=0.41\tw,bb=128 369 592 715,clip=true]{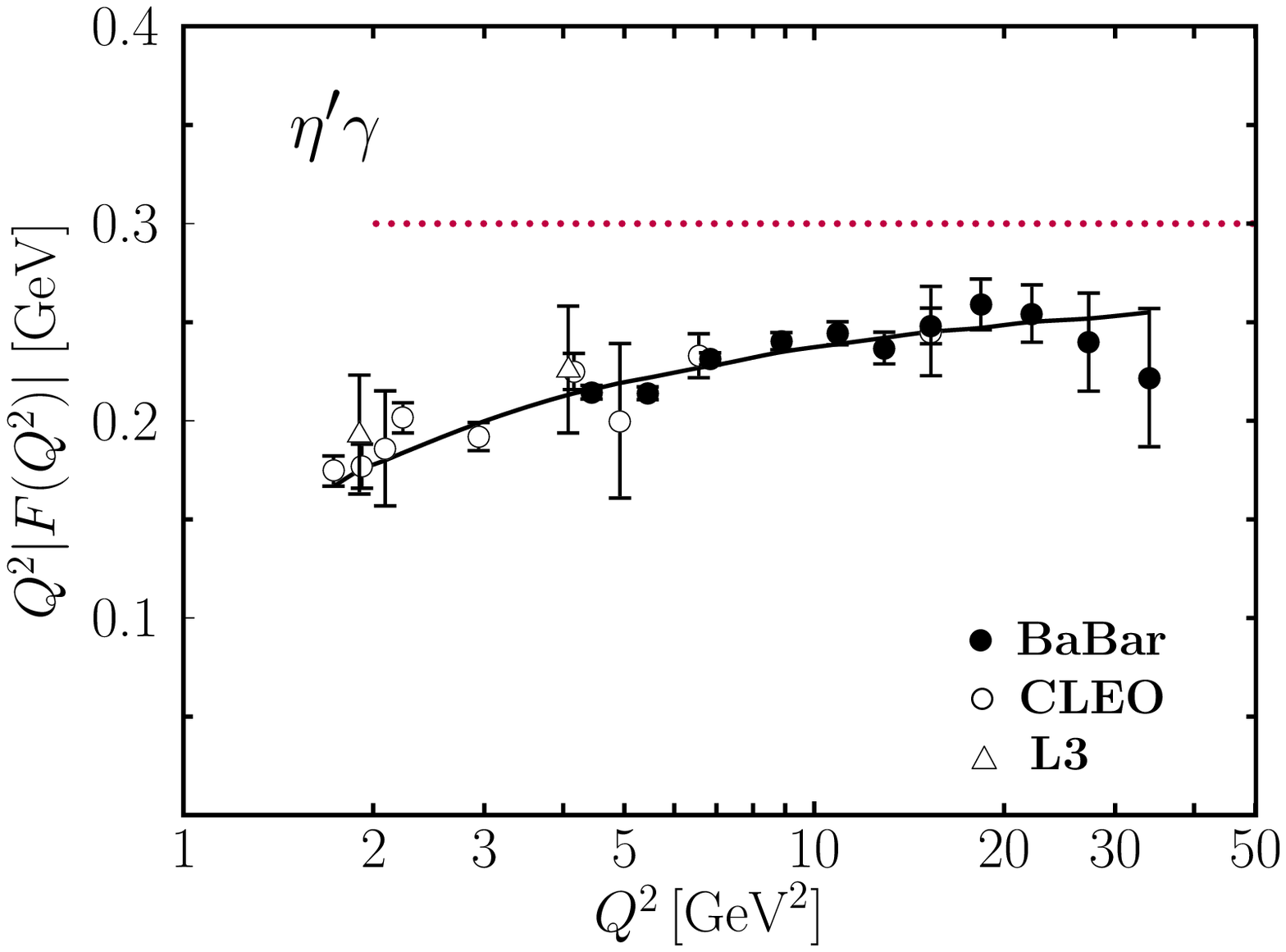}
\end{center}
\caption{The $\eta\gamma$ (left) and $\etap\gamma$ (right) transition form
  factors scaled by $Q^2$. Data taken from \ci{babar,cleo,l3}. The solid lines
  represents our default fit (see Tab.\ \ref{tab:fit}). The dotted lines are 
  the LO asymptotic results \req{eq:asymp}.}
\label{fig:eta}
\end{figure}

Using this standard setting together with the mixing parameters derived in
\ci{FKS1}, we fit the Gegenbauer coefficients of order 2 to the
data~\footnote{
The signs of the transition form factors are not measured.}~\footnote
{The CLEO collaboration \ci{cleo} measured the form factors  for various $\eta
(\etap)$ decay channels. We take into account all of them. In cases where for
a given value of $Q^2$ there are several values of the form factor we use their
error-weighted average.}~\footnote
{Above $Q^2=2\,\gev^2$ there is only one $\etap\gamma$ data point from \ci{l3}
with a large error. It has no bearing on our fits and will therefore 
be not mentioned explicitly in the following.} 
\ci{babar,cleo}. The result of this fit, termed the default fit in the
following, is quoted in Tab.\ \ref{tab:fit} and shown in Fig.\ \ref{fig:eta}. 
For comparison the results of fits to only the CLEO data \ci{cleo} and 
only the BaBar data \ci{babar} are also shown in Tab.\ \ref{tab:fit}. 
The latter fit can be regarded as a change of the minimal value of $Q^2$ 
for which data are taken into account in the fits. 

One notices that the 
three sets of parameters for the quark \da s agree quite well with each
other. Deviations of a little more than $1\sigma$ are seen for $a_2^g$. The 
effect of the Babar data in combination with the CLEO one results in a 
reduction of the parameter errors and a more precise determination of
$a_2^g$. In contrast to our previous work where we have had at disposal only 
the CLEO data \ci{cleo} and have chosen $\muR=Q/\sqrt{2}$ and $\muF=Q$, the 
gluon \da{} is not compatible with zero now. The reasonable agreement of the 
three fits demonstrates the consistency of the CLEO \ci{cleo} and BaBar data 
\ci{babar}. The $\chi^2$ of the default fit is very good given that all
together 40 data points are included in the fit. In Fig.\ \ref{fig:eta} the 
fit is compared to experiment. 

\begin{figure}[t]
\begin{center}
\includegraphics[width=0.31\tw]{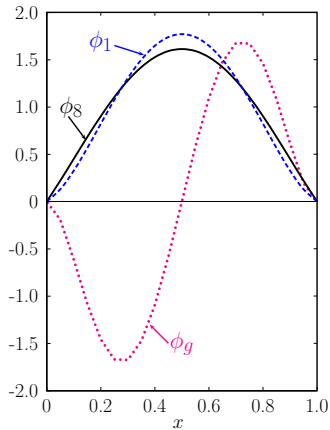}
\end{center}
\caption{The \da s specified by the default fit at the initial scale.}
\label{fig:da}
\end{figure}

The \da s corresponding to the default fit (see Tab.\ \ref{tab:fit}) are 
shown in Fig.\ \ref{fig:da}. Those for the octet and singlet $\qbq$ components 
are close to the asymptotic form of a meson \da{}. They are symmetric around 
$x=1/2$ while the gluon \da{} is antisymmetric. We repeat that the \da s are 
to be considered as effective ones since the parameters $a_2$ are contaminated 
by higher order Gegenbauer coefficients. 

\subsubsection{Scale dependence}
\label{sec:scale}
Let us discuss the scales dependence in more detail. 
In principle, renormalization and factorization scales 
can be chosen independently. 
In this work we have chosen $\muR=\muF=Q$ for the default fit
but we also investigate other choices of the scales in order to
learn about the theoretical uncertainties of our results.

For the renormalization scale the usual choice is $\muR=Q$.
On the basis of a next-next-to-leading order calculation 
of the pion transition form factor it has been argued 
in \ci{MelicNP01,MelicNP01a} that another possible
choice of $\muR$ is the square root of the average virtuality instead 
of the characteristic scale of the process, $Q$. For the transition form 
factors the average virtuality
is $Q^2/2$. Hence, in order to explore the renormalization 
scale dependence of the resulting \da s we also utilize this choice. 
As an inspection of Tab.\ \ref{tab:fit} reveals the dependence of the
fit on the renormalization scale is rather strong; the 
change of $\muR$ from $Q$ to $Q/\sqrt{2}$ results in substantial change 
of the parameters although both the fits are of similar quality.  
Thus, the theoretical uncertainties of our results are larger than indicated
by the errors of the fitted Gegenbauer coefficients. 

It is important to note that our NLO calculation is 
in fact the LO calculation in $\als$ 
and it is a well known fact that in order to stabilize 
the dependence on $\muR$ the NLO QCD corrections 
i.e., NNLO corrections to the transition form factor, should be included. 
At NNLO the presence of $\als^2 \ln \muR^2/Q^2$ terms
stabilizes the dependence on $\muR$ and all predictions fall
relatively close 
(see, for example, \cite{MelicNP98} for discussion on that point). 
Concerning fits, one would thus expect that the inclusion 
of NNLO would considerably decrease the variation
of the obtained Gegenbauer coefficients with $\muR$. 
Without this stabilizing effect of NNLO we are left with the variation
illustrated in Tab.\ \ref{tab:fit}.
We note that in order to circumvent this scale ambiguity
one can claim to have found
a sensible renormalization scale setting
a number of which has been proposed in the literature
(see Ref.  \cite{Wu:2013ei} and references therein) as, for example,
the principle of maximum conformality 
where all non-conformal terms associated with the $\beta$-function
in the perturbative series are summed into the running coupling
and a scale-fixed prediction is obtained.

Next, we turn to the factorization scale choice.
As elaborated in \cite{MelicNP01a}, the dependence on $\muF$
can be cancelled order by order in $\als$
by performing the resummation of $(\als \ln \muF^2/Q^2)^n$ terms
up to the characteristic scale of the (one-scale) process $Q$,
and this turns out to be an equivalent of the choice $\muF=Q$.
Still, to test the strength of the residual dependence 
of the expression \req{eq:FFexpansion} on the factorization scale 
we perform a fit with the choice $\muR=\muF=Q/\sqrt{2}$.
From the results, quoted in Tab.\ \ref{tab:fit}, one sees that
the fit mildly depends on $\muF$. The quark Gegenbauer coefficients 
hardly change. For $a_2^g$ the factorization scale dependence is a bit 
more pronounced although the values of $a_2^g$ agree within errors  
for the two fits with different $\muF$ but the same renormalization scale.

\subsubsection{Parameter correlations  and evolution of Gegenbauer
coefficients}
\label{sec:corr}
\begin{table*}[t]
\renewcommand{\arraystretch}{1.4} 
\begin{center}
\begin{tabular}{|c || c | c | c | }
\hline     
Scales &  $a_2^1 - a_2^g$  & $a_2^8 - a_2^1$  & $a_2^8 - a_2^g$   
\\[0.2em]\hline
$\mu_R^2=\mu_F^2=Q^2$ &  0.371 & 0.248 & 0.057 \\[0.2em]   
$\mu_R^2=Q^2/2$ &  0.607 & 0.228 & 0.057 \\[0.2em]  
$\mu_R^2=\mu_F^2=Q^2/2$ &   0.760 & 0.202 & 0.058 \\[0.2em]
\hline
\end{tabular}
\end{center}
\caption{Correlation coefficients obtained in the fits to the CLEO 
\ci{cleo} and BaBar \ci{babar} data for various choices of the scales.
Except of the scales the standard setting and the mixing parameters 
  of \ci{FKS1} are used.}
\label{tab:correlation}
\end{table*}

The parameter correlations
offer an additional
quantitative and qualitative
insight into our fits 
so we present them here,
as well as, comment on 
the general behavior
of the Gegenbauer coefficients
under evolution.

\begin{figure}[t]
\begin{center}
\includegraphics[width=0.43\tw]{fig7-corrQG.eps}\hspace*{0.03\tw}
\includegraphics[width=0.40\tw]{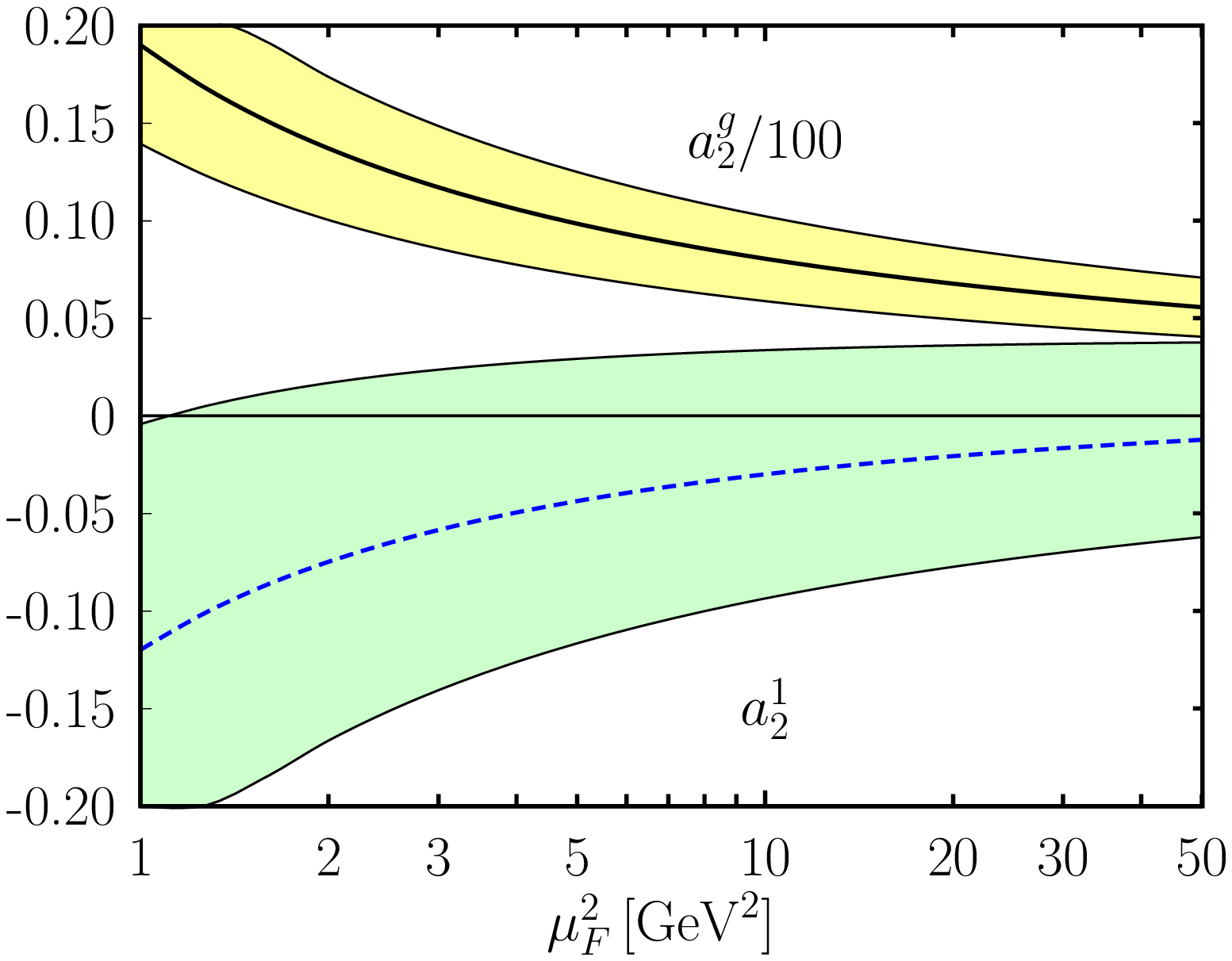}
\caption{Correlation between $a_2^1$ and $a_2^g$ (left) and their evolution
  (right) for the standard setting. The $1\sigma$ $\chi^2$-contours are shown 
  for fits to the data from only CLEO \ci{cleo}, only BaBar \ci{babar} and to 
  both data sets. The shaded bands indicate the errors of the coefficients.}
\label{fig:contour}
\end{center}
\end{figure}

Due to mixing under evolution we expect the strongest correlation
between the Gegenbauer coefficients $a_2^1$ and $a_2^g$. Inspection of Fig.\ 
\ref{fig:contour} reveals that the correlation is particularly strong for the 
fit to just the CLEO data but becomes milder for the fit to the combined CLEO 
and BaBar data. This goes parallel with a reduction of the parameter errors. 
The strength of the correlation between $a_2^1$ and $a_2^g$ 
depends on the chosen factorization and renormalization scales; it is smallest
for the standard setting, see Tab.\ \ref{tab:correlation} where we compile the
correlation coefficients for the three fits to the CLEO and BaBar data.
So the higher the scales, the lower are the correlations between
$a_2^1$ and $a_2^g$.
Furthermore, the correlation between $a_2^1$ and $a^8_2$ is mild while 
$a_2^g$ and $a_2^8$ are nearly uncorrelated. The origin of the correlation 
between $a_2^8$ and the other Gegenbauer coefficients lies in $\eta - \etap$
mixing. The form factor $F_{\eta\gamma}$ is dominated by the octet
contribution ($a_2^8$) while $F_{\etap\gamma}$ is mainly fed by the 
singlet one ($a_2^1, a_2^g$). 

The evolution of the Gegenbauer coefficients $a_2^1$ and $a_2^g$ is shown on
the right hand side of Fig.\ \ref{fig:contour}. The coefficients decrease
relatively fast from the initial scale, up to about $10$ GeV$^2$ and after
that the approach to zero is slow. This behavior is a consequence of 
the properties of the logarithm contained in $\als(\muF)$ (see \req{eq:ang}): 
for $\muF\gg \muO$ (in fact for $\muF>3\,\gev$) the derivative of the
logarithm is very small. The Gegenbauer coefficients of the octet \da{} evolve 
similarly. This flat behavior of the logarithm for large scales 
is also partially responsible for the fact that even in a NLO calculation one 
can determine only one Gegenbauer coefficient for each \da{} or, more 
precisely, one linear combination of them. 

 \subsubsection{OZI rule constraints}

The
opposite Fock components from Eqs. (\ref{qsfock}-\ref{qsfock2})
lead to violations of the OZI rule if they were
not suppressed. 
Hence, 
one expects that
\be
\frac{\mid \phi_{\rm opp}(x,\muF)\mid}{\phi_{\rm AS}(x)} \ll 1
\label{eq:requirement}
\ee
holds for any values of $x$ at least for a limited range of the factorization
scale. Asymptotically, where all \da s evolve in $\phi_{\rm AS}$, $\phi_{\rm opp}$ 
is zero anyway. To a sufficient degree of accuracy \req{eq:requirement} can be 
replaced by
\be
 \frac{\sqrt{2}}3\,|a_2^1-a_2^8| \ll 1
\label{eq:requirement2}
\ee
Indeed, as can be seen from Table \ref{tab:fit},
the default fit meets \req{eq:requirement2} at the initial scale 
and, as can readily be checked, at all larger factorization scales: 
$\frac{\sqrt{2}}3\,|a_2^1-a_2^8| \lsim 0.03$. Hence, no substantial
violations of the OZI rule follow from the \da s specified by the
default fit and shown in Figs.\ \ref{fig:da}. Also the fits using different
choices of $\muR$ and $\muF$ as well as those to just the CLEO or BaBar data
satisfy \req{eq:requirement2}. 

In order to avoid large violations of the OZI rule one may follow a suggestion 
made in \ci{ball07} and assume $a_2^1\equiv a_2^8$ at the initial scale. 
Although evolution generates some violations of the OZI rule with increasing 
scale they are always sufficiently small. The fit assuming $a_2^1\equiv a_2^8$
and using the mixing parameters of \ci{FKS1,FKS2} is still of reasonable
quality. Although $\chi^2$ is somewhat larger, the fit parameters are similar 
to those obtained from the three-parameter fit, see Tab.\ \ref{tab:fit}. The 
difficulties with the OZI rule of the mixing parameters determined in 
\ci{escribano05} (see Tab.\ \ref{tab:mix})
is corroborated by the analogous fit with 
$a_2^1(\muO)\equiv a_2^8(\muO)$. This fit fails badly, $\chi^2$ is 240. 

Let us end with some additional comments on various mixing
parameters from the literature in the context of OZI rule
violations.
Using the $\eta-\etap$ mixing parameters determined in \ci{escribano05}
which markedly differ from those given in \ci{FKS1,FKS2}, 
one also arrives at a reasonable fit to the form factor data with regard to 
$\chi^2$. However, in this case $a_2^8$ is positive leading to rather large
OZI rule violations. The mixing parameters given in \ci{escribano05} are 
therefore to be employed with reservation. It is to be mentioned that the work 
\ci{escribano05} has also been criticized in \ci{Klopot:2011qq} on the basis
of an analysis of the transition form factors with a $U_A(1)$-anomaly sum rule. 
The theoretical set of mixing parameters discussed in \ci{Kroll05} is
intermediate between \ci{FKS1,FKS2} and \ci{escribano05}. The quality of the 
fit to the form factor data is similar to other fits but the difference
$|a_2^1-a_2^8|$ is rather large although smaller than for the fit using the 
mixing parameters given in \ci{escribano05}. Thus, with regard to the strength 
of OZI rule violation the mixing parameters given in \ci{FKS1,FKS2} seem to be 
favored. Ideally one should fit the mixing parameters together with the lowest 
Gegenbauer coefficients to the data. Unfortunately such a multi-parameter fit 
does not lead to a reasonable solution, there are extremely strong
correlations among the parameters, often large violations of the OZI rule and
a covariance matrix that is not positive  definite. Thus, we refrain from
discussing such fits.

\subsection{Comparison to other results}
\label{sec:comparison}

We here briefly discuss our results in relation
to some relevant results found in the literature.

Ali and Parkhomenko \ci{ali03} have performed an analysis of the $\etap$
energy spectrum in the inclusive decay 
$\Upsilon(1S)\to ggg^* \to\etap X$ \ci{cleo02} in order to
constrain the $\etap$-meson \da. At an intermediate step of the analysis
of the $\Upsilon(1S)\to\etap X$ energy spectrum the $g^*\to \etap g$
transition form factor is to be calculated (see Sect.\ 6). From the high end
of the $\etap$ meson energy spectrum Ali and Parkhomenko determine
the Gegenbauer coefficients $a_2^1$ and $a_2^g$ which are in agreement
with our results within very large errors. In order to calculate the
energy spectrum also at low and even negative gluon virtualities Ali and 
Parkhomenko introduce a positivity constraint for the $\etap g$ transition 
form factor which is achieved by choosing $\muF^2=\muR^2=|Q^2|+m^2_{\etap}$.
This constraint significantly reduce the allowed range for the Gegenbauer
coefficients and, in combination with our previous result \ci{gluon}
they obtain
\be
a_2^1\= -0.08\pm 0.03\,,  \qquad a_2^g\=6.5\pm 2.7\,.
\ee
at the initial scale $\muO=1\,\gev$. While the value for $a_2^1$
agrees with our default result within errors, our result for the coefficient
$a_2^g$ is larger as a consequence of the BaBar data which were
not available to the authors of \ci{ali03}. The role of the positivity
constraint used in \ci{ali03} remains to be understood.

\begin{figure}[t]
\begin{center}
\includegraphics[width=0.46\tw,bb=121 371 585 716,clip=true]{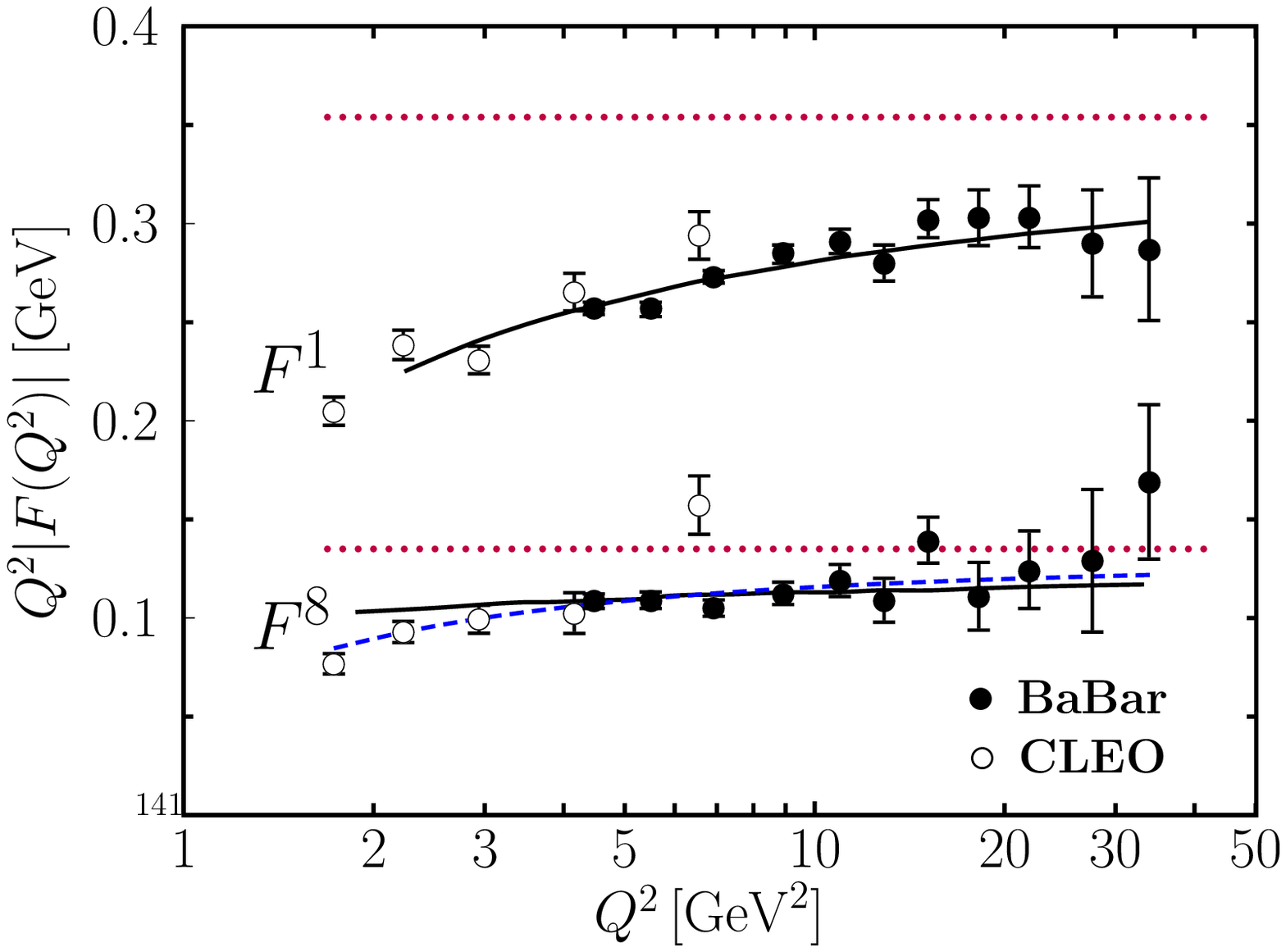}
\end{center}
\caption{The octet and singlet components of the transition form factors
  evaluated with the mixing parameters given in \ci{FKS1,FKS2}.
 Data taken from \ci{babar,cleo}. The solid and dotted lines represent our 
 default fit (see Tab.\ \ref{tab:fit}) and the asymptotic results, 
 respectively. The dashed line is the fit with a power correction 
 ($a_2^8=0.06\pm 0.05$, $c_8=-0.04\pm 0.02$).}
\label{fig:octet}
\end{figure}

The lowest Gegenbauer coefficient of the octet \da{} has been calculated with
the help of QCD sum rules in Ref. \ci{ball98}. A value of about $0.2$ with a large
uncertainty has been obtained for the coefficient $a_2^8$. In this calculation
the \da{} is normalized to $f_\eta^8=f_\pi$. For the larger octet decay
constant (see Tab.\ \ref{tab:mix}) we are using the value of $a_2^8$ is
expected to be somewhat smaller, say, about 0.16. Given the large  uncertainty
of the sum rule result this value is not in conflict with our default fit. 
However this large face value of the sum rule result can only be reconciled with 
the form factor data if one allows for an additional power correction (or for 
higher Gegenbauer coefficients) which compensate the positive $a_2^8$ to a
large extent. In order to examine this possibility we extract 
the octet contribution to the transition form factors using \req{eq:ossum},
 and \req{mix81}
\be
F_{\eta_8\gamma}\=\frac{\cos{\theta_1}
                    F_{\eta\gamma}+\sin{\theta_1}F_{\etap\gamma}}
                    {\cos{(\theta_8-\theta_1)}}
\ee
and perform a fit analogously to the ones described above but allowing for an
additional power correction $c_8/Q^4$ to it. Keeping $a_2^8=0.16$ fixed 
and using the mixing parameters determined in \ci{FKS1,FKS2}, we obtain 
$c_8=-0.08\pm 0.01$ from the fit to the data on $F_{\eta_8\gamma}$. For the 15 
data points the minimum $\chi^2$ is 21 which is somewhat larger than our best 
result ($\chi^2=15.3$ for $a_2^8=-0.01\pm 0.02$, $c_8=0$) but still
tolerable. Freeing also $a_2^8$ a very good fit is obtained ($a_2^8=0.06\pm
0.05$, $c_8=-0.04\pm 0.02$, $\chi^2=16.3$). The latter fit is shown in Fig.\ 
\ref{fig:octet} and compared to experiment. In this figure also the results of
the default fit for the octet and singlet form factors 
\be
F_{\eta_1\gamma}=\frac{\cos{\theta_8}
                    F_{\etap\gamma}-\sin{\theta_8}F_{\eta\gamma}}
                    {\cos{(\theta_8-\theta_1)}}
\ee
are displayed and compared with experiment.

\subsection{A comment on the time-like data}
The BaBar collaboration 
\ci{babar06} has measured the $\eta\gamma$ and $\etap\gamma$ transition form
factors at $s=112\,\gev^2$. We do not include these data in our analysis since 
the theoretical treatment of form factors in the time-like region
is non-trivial and not well understood. In the time-like region the form
factors are no longer real and there are subtleties regarding the analytic
continuation from the space- to the time-like region \ci{Bakulev:2000uh} (and
references therein). 

Compared with the naive expectation that at $s=112\,\gev^2$ the form factor
should be close to the asymptotic prediction the Babar result for the
$\eta\gamma$ form factor is about $2\sigma$ too large.
This discrepancy is likely not a consequence of the description of $\eta-\etap$
mixing. The quark-flavor mixing scheme is on sound theoretical grounds and is 
phenomenologically well established in a large variety of processes. A
$2\sigma$ discrepancy for a single data point cannot discard this mixing
scheme. In contrast to the case of the $\eta$ the $\etap\gamma$ form
factor at $s=112\,\gev^2$ measured by the BaBar collaboration \ci{babar06}, 
is in agreement with the naive expectation.

\subsection{The $g^*g^*P$ vertex}

We are now in the position to repeat the evaluation of the $g^*g^*P$ form 
factors we performed in \ci{gluon}. With the results on the gluon \da{} 
obtained from the present analysis of the $P\gamma$ transition form factors
we believe to have more precise results for the $g^*g^*P$ form factors
now. On the importance of these form factors we have already commented in 
the introduction. 

The gluonic vertex is written analogously to the electromagnetic one as
\begin{equation}
    \Gamma^{\mu \nu}_{a b}=
  i  \; F_{Pg^*}(\ov{Q}^{\,2},\omega) 
    \; \delta_{ab} \; \eps^{\mu \nu \alpha \beta} 
    \; q_{1\alpha} q_{2\beta}\,,
\label{eq:Gmunuab}
\end{equation}
where $q_1$ and $q_2$ now denote the momenta of the gluons and $a$ and $b$
label the color of the gluon. 

\begin{figure}[t]
\begin{center}
\includegraphics[width=0.65\tw]{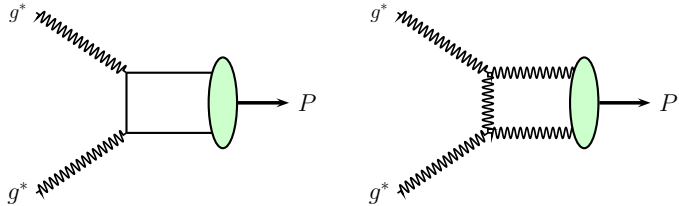}
\end{center}
\caption{ Sample lowest order Feynman graphs for the $g^*g^*\to q\bar{q}$ 
and  $g^*g^*\to gg$ subprocesses to the $g^*g^*\to P$ transition form factors.}
\label{fig:gg-graph}
\end{figure}
According to \ci{gluon} the $Pg^*$ transition form factor to leading-twist
accuracy and lowest order of $\als$ is to be calculated from Feynman graphs 
of which examples are shown in Fig. \ref{fig:gg-graph}. 

\begin{figure}[t]
\begin{center}
\includegraphics[width=0.45\tw]{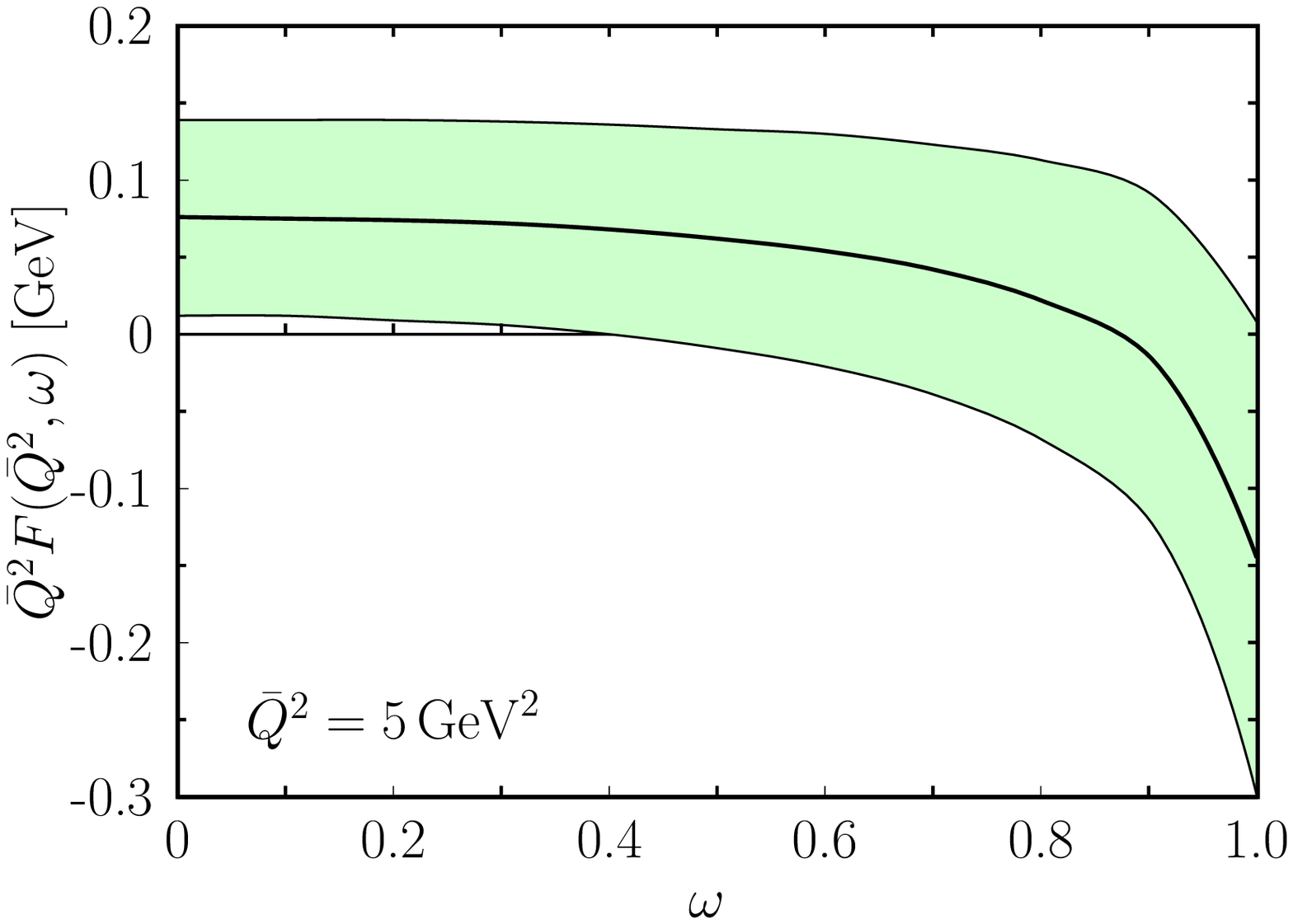}
\includegraphics[width=0.45\tw]{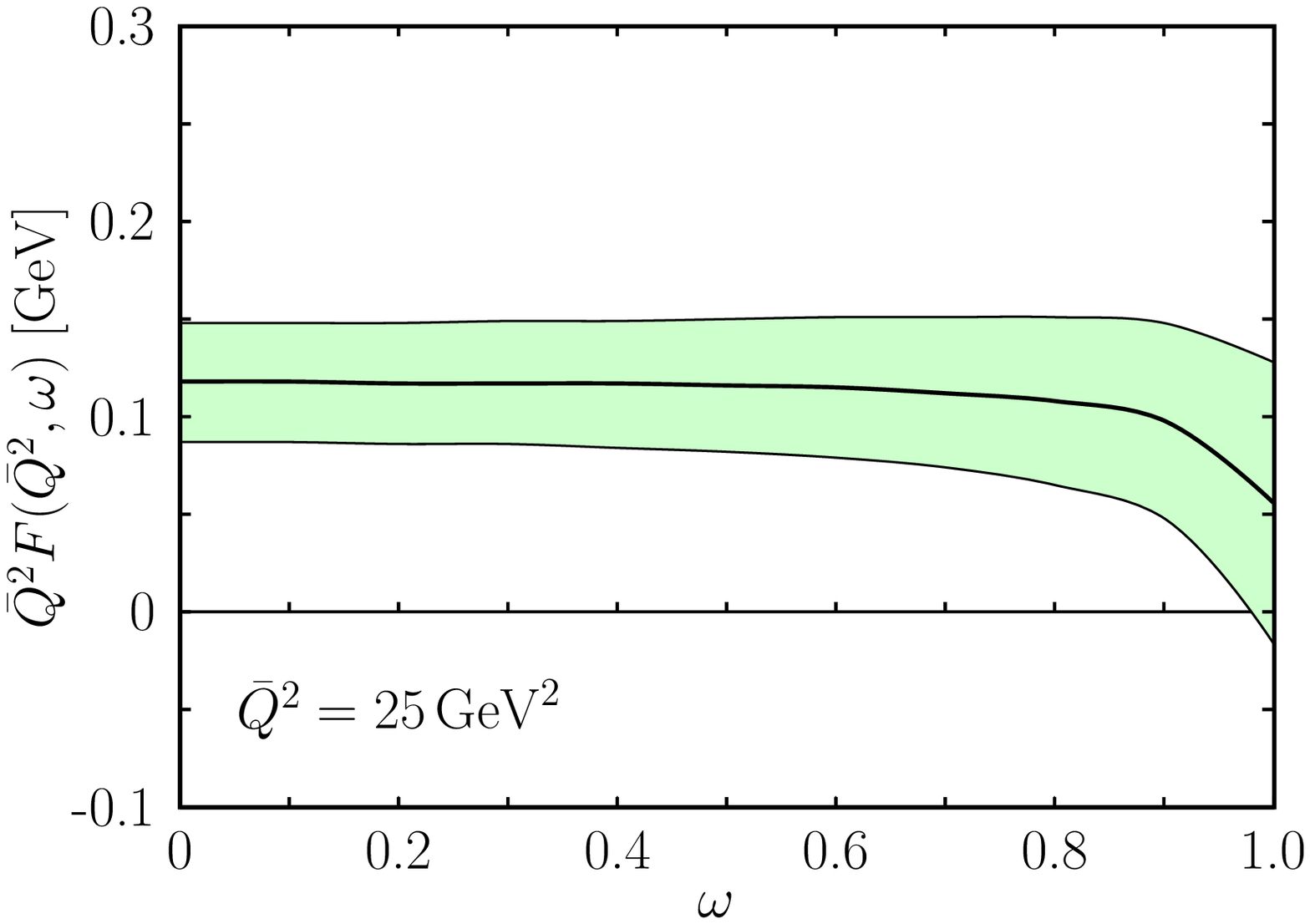}
\end{center}
\caption{Predictions for the $g^*g^*\etap$ form factor evaluated from
the default solution with $\muR^2=\muF^2=\bar{Q}^{\,2}$. The shaded bands
indicate the $1\sigma$ uncertainty of the predictions.}
\label{fig:ggetap}
\end{figure}

The results are given in \ci{gluon} and are not repeated here. 
Using the Gegenbauer coefficients of the default fit, we
can readily evaluate the $g^*g^*\to \etap$ transition form factor. The results,
including the $1\sigma$ error band, are shown in Fig.\ \ref{fig:ggetap} for
two values of $\bar{Q}^{\,2}=\muR^2=\muF^2$. As compared to our previous
results \ci{gluon} the error bands are markedly narrower while the central
values do not differ much.

As for the $\eta\gamma$ and $\etap\gamma$ transition form factors and in
order to be consistent with that analysis power corrections are neglected
here as well. For the $g^*g^*\eta (\etap)$ vertex function, in particular
for $\omega\to 0$, any power corrections as for instance quark transverse
momenta or meson mass corrections, are small since the vertex function is
not end-point sensitive in this kinematic limit. Mass corrections to the 
$g^*g^*\etap$ vertex function have been estimated by Ali and Parkhomenko 
\ci{ali03,ali2-03}.

\section{The $\pi\gamma$ transition form factor }
Finally, we present the preliminary analysis and discussion
of the $\pi\gamma$ transition form factor
which is inspired by the availability of both
\ci{babar} and \ci{BELLE12} data.

\begin{figure}[t]
\begin{center}
\includegraphics[width=0.45\tw,bb= 107 384 567 722,clip=true]
{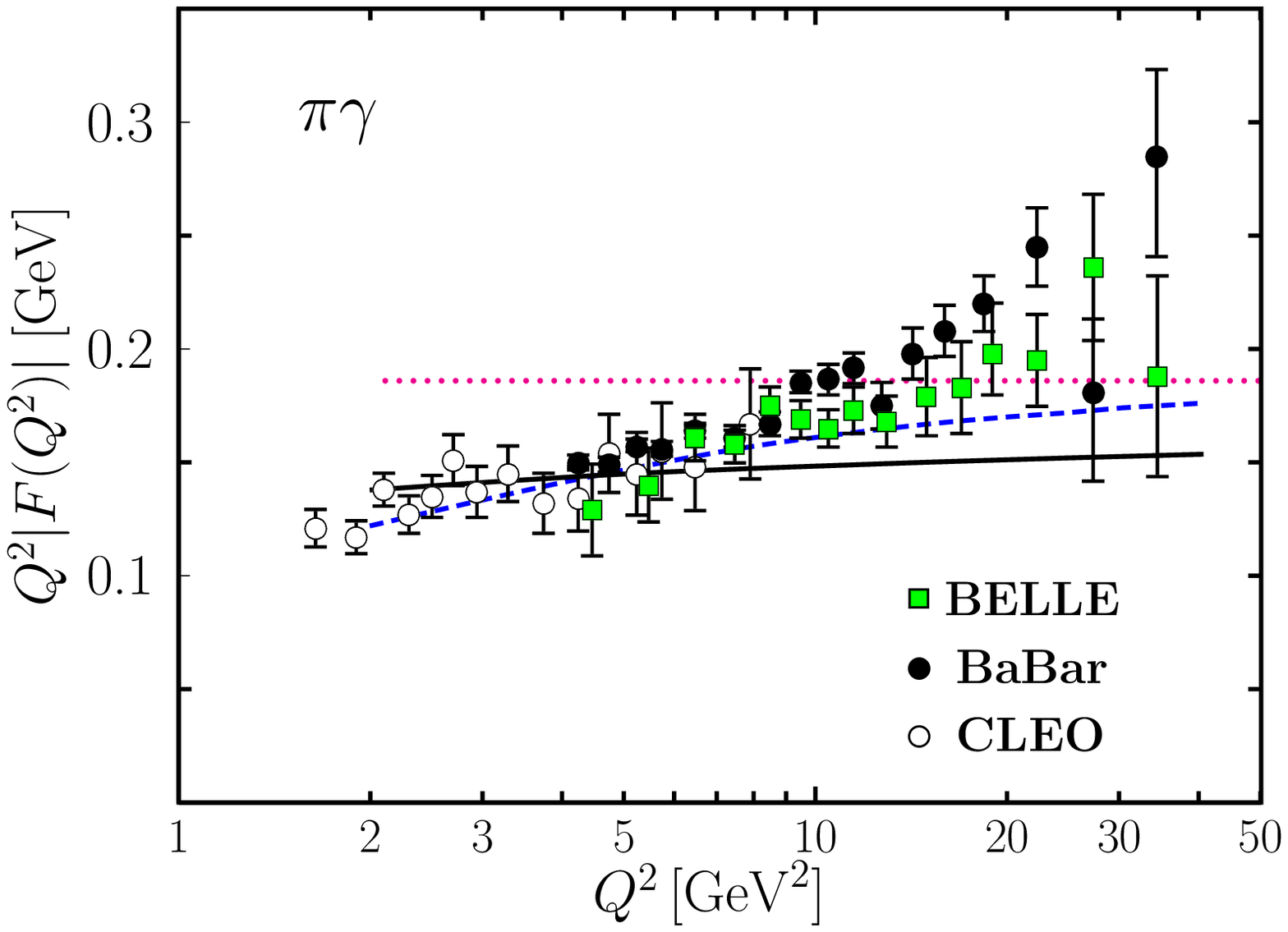}
\end{center}
\caption{The $\pi\gamma$ transition from factor. Data taken from
\ci{babar,BELLE12,cleo}. The solid (dashed, dotted) line represents
a fit with $a_2^\pi(\muO)=-0.02$ (\ci{KrollR96}, LO asymptotic result).}
\label{fig:piga}
\end{figure}

For comparison we show in Fig.\ \ref{fig:piga} the data on the $\pi\gamma$
transition form factor \ci{babar,BELLE12,cleo}.
It is important to realize the dramatic difference between the BaBar data
\ci{babar} on the $\pi\gamma$ form factor and the $\eta (\etap)\gamma$ data.
In contrast to the latter a leading-twist analysis of the $\pi\gamma$ 
form factor to fixed order of perturbative QCD fails because the data do not 
seem to respect dimensional scaling. Power corrections seem to be demanded 
by the $\pi\gamma$ data as well as a positive value of the 2nd Gegenbauer 
coefficient of the pion \da{}, see for instance \ci{piga11,stech,agaev11}.
This difference between the $\pi\gamma$ and the other form factors implies 
a strong breaking of flavor symmetry which
has never been observed in the sector of pseudoscalar mesons before.  

On the other hand, the BELLE data \ci{BELLE12} do not show this sharp rise
with $Q^2$, they lie systematically below the BaBar data. They do not
exceed the asymptotic result for the $\pi\gamma$ form factor and are close
to the behavior of the NLO leading-twist approach as is the case of the other
transition form factors. An example of such a fit to the data from \ci{cleo}
and \ci{BELLE12} is shown in Fig.\
\ref{fig:piga} (with $a_2^\pi(\muO)=-0.02\pm 0.02$ and $\chi^2=34.9$ for 
28 data points). Although the $\chi^2$ of this fit is reasonable the $Q^2$
dependence of the fit seems to be too flat as compared to the BELLE data
which may be viewed as a hint at lacking power corrections. For instance,
a result obtained with $k_\perp$ factorization which encodes corrections
of order $\langle k^2_\perp/Q^2\rangle$, is in good agreement with the 
CLEO and 
BELLE data \ci{KrollR96}. We finally emphasize that the BELLE data 
imply only mild violations of flavor symmetry. A more detailed comparison
of the BaBar and BELLE data with theoretical models can be found in
\ci{stefanis}. Although the BELLE data seem to be favored against the BaBar
data with regard to the standard theoretical concepts, an understanding
of the origin of the discrepancy between the two measurements is required. 

\section{Summary}

We have analyzed the $\eta\gamma$ and $\etap\gamma$ transition form factors 
within a collinear factorization approach to leading-twist accuracy and
NLO of perturbative QCD. The analysis of the data \ci{babar,cleo}
allowed for an extraction of the lowest 
Gegenbauer coefficients of the $\qbq$ flavor-octet and singlet \da s as well
as that of the glue-glue \da. The Gegenbauer coefficients are better
determined now and have smaller errors than those extracted in 
\ci{gluon}. Our default result for $\muO=1\,\gev$ is (see Tab.\ \ref{tab:fit}):
\be
a_2^8\=-0.05\pm 0.02\,, \qquad a_2^1\=-0.12\pm 0.01\,, \qquad a_2^g\=19\pm
5\,.
\ee
There are a number of uncertainties of this result. First there is the
uncertainty due to the chosen renormalization scale. 
Next, despite the large range of $Q^2$ covered by data now one still cannot 
determine more than the lowest Gegenbauer coefficients of the three \da s. 
The coefficients we quote are to be regarded as effective parameters which are
contaminated by higher order Gegenbauer coefficients. In fact, according to
the discussion at the end of Sect.\ \ref{sec:corr}, it seems impossible to
extract more information from the transition form factors than the lowest
Gegenbauer coefficients. In order to determine more Gegenbauer coefficients
additional processes have to be analyzed. Finally, given the quality of the 
present data, it is still not possible to discriminate between the logarithmic 
$Q^2$ dependence generated by the evolution and the NLO corrections,
and power corrections (see the
discussion in Sect.\ \ref{sec:comparison}). We have neglected power correction 
in our analysis. Allowing for power corrections in the form factor analysis one
may for instance find positive $q\bar{q}$ Gegenbauer coefficients as is the
case for the pion \da{} (e.g.\ \ci{agaev11}).
  
As an application of our results for the Gegenbauer coefficients we calculated
the $g^*g^*\etap$ vertex function.
Another processes in which the gluon-gluon Fock components of the $\eta$ and
$\eta^\prime$ mesons may play an important role, are the $\chi_{cJ}$ ($J=0,2$)
decays into pairs of $\eta$ and/or $\etap$ mesons. Here, the $c$-quark 
mass is considered to be large enough to allow for a perturbative treatment of
these decays. An explicit calculation of the perturbative contribution to the 
$\chi_{cJ}$  decays taking into account the two-gluon Fock components, is
tedious, many Feynman graphs contribute even to lowest order \ci{bai85}.
Moreover, there is another complication. As is shown in \ci{BKS2} the next 
higher Fock state, $\cbc g$, of the $\chi_{cJ}$, the so-called color-octet 
contribution \ci{braaten95}, is also to be taken into account since it scales 
with the same power of the hard scale, $m_c$, as the $c\bar{c}$ contribution.
With regard to these complications whose detailed calculation 
is very time-consuming, we will not attempt a complete analysis of these decay
processes; this is beyond the scope of the present paper. A statement whether
or not our $gg$ \da s are in conflict with the peculiar features of the
$\chi_{cJ}$ decays \ci{thomas07,pdg,bes3} is therefore premature.

This work was supported in part by the BMBF under the contract No.\ 06RY9191
and in part by Croatian Ministry  of Science, Education and Sport
under the contract no. 098-0982930-2864.

\end{document}